\documentclass[reqno]{amsart}
\usepackage{amsaddr}
\usepackage{graphicx,amsmath}
\usepackage{float}
\usepackage[utf8]{inputenc}
\usepackage{color}
\usepackage{cite}
\usepackage{fancyhdr}
\usepackage[margin=2cm]{geometry}
\usepackage{appendix}
\usepackage{multirow}
\usepackage[justification=centering]{caption}
\floatstyle{plaintop}
\restylefloat{table}

\begin{document}
\title{Challenges in the thermal modeling of highly porous carbon foams}

\author{A. Fehér$^{1,5}$, R. Kovács$^{1,2,5}$, Á. Sudár$^{2,3,4}$, G. G. Barnaföldi$^{2}$}

\address{
$^1$Department of Energy Engineering, Faculty of Mechanical Engineering, BME, Budapest, Hungary \\
$^2$Wigner Research Centre for Physics, Institute for Particle and Nuclear Physics, Budapest, Hungary \\
$^3$Department of Diagnostic Radiology, National Institute of Oncology, Budapest, Hungary \\
$^4$Institute of Nuclear Techniques, Faculty of Natural Science, Budapest University of Technology and Economics, Budapest, Hungary\\
$^5$Montavid Thermodynamic Research Group
}

\date{\today}

\begin{abstract}
The heat pulse (flash) experiment is a well-known, widely used method to determine thermal diffusivity. However, for heterogeneous, highly porous materials, neither the measurement nor the evaluation methodologies are straightforward. In the present paper, we focus on two open-cell carbon foam types, differing in their porosity but having the same sample size. Recent experiments showed that a non-Fourier behaviour, called 'over-diffusive' propagation, can be present for such a complex structure. The (continuum) Guyer\,--\,Krumhansl equation stands as a promising candidate to model such transient thermal behaviour. In order to obtain a reliable evaluation and thus reliable thermal parameters, we utilize a novel, state-of-the-art evaluation procedure developed recently using an analytical solution of the Guyer\,--\,Krumhansl equation. Based on our observations, it turned out that the presence of high porosity alone is necessary but not satisfactory for non-Fourier behaviour. Additionally, the mentioned non-Fourier effects are porosity-dependent, however, porous samples can also follow the Fourier law on a particular time scale. These data serve as a basis to properly identify the characteristic heat transfer mechanisms and their corresponding time scales, which altogether result in the present non-Fourier behaviour. Based on these, we determined the validity region of Fourier's law in respect of time scales. 


\end{abstract}
\maketitle

\section{Introduction} 
Together with the development of advanced manufacturing technologies, materials with complex inner structures appeared in the engineering practice. The one we place the focus on is an open-cell, highly porous carbon foam material, which has particular mechanical and thermal properties. Its lightweight, highly porous structure promotes that material to be exceptional in applications requiring large surface/volume (or mass) ratio such as supercapacitors~\cite{ZhangEtal18} and chemical synthesis~\cite{YangEtal20}. Besides, foams are also outstandingly advantageous for particle detectors, especially in large-scale detectors, where the low material budget is a key factor. In that case, less mass density means much less probability of unwanted particle scattering, and that greatly increases the reliability and accuracy of sensors being attached onto that structure. As these semiconductors also dissipate power, they need cooling, which can be more efficient thanks to the open-cell design of the foam while having acceptable mechanical stability. In the following, we provide a brief introduction about the specific aspects we focus on in the present study.

\subsection{Low material budget detectors} 
One of the novelties of recent detector developments in High Energy Physics (HEP) is to build detectors with as low material budget as possible. A lightweight detector setup absorbs fewer particles, therefore, it can measure more details of the elementary processes. This novel R\&D direction is presented by the application of the thin silicon pixel detectors, where the thickness of a sensor layer is getting close to the $\mathcal{O}(50~\mu$m) size~\cite{Mager:2016}. Building such a detector required large and solid support frames, with minimal material budget as well. For this aim, the application of light material foams can be an excellent option, especially since porosity offers good options for air cooling as well.

As an example, one can see the ALICE detector upgrade plans at the CERN's Large Hadron Collider~\cite{ALICEWeb}. The upgrade of the Inner Tracking System (ITS3) is planned for the Long Shutdown 3 (LS3) during the period 2025-2027~\cite{ITS3:2019,Suljic:2020}. This development will include a new vertex tracker based on truly cylindrical wafer-scale semiconductor sensors  with a material budget, $<0.05\% $~X/X$_0$ per layer, and located as close as 18~mm to the interaction point~\cite{Domenico:2022,Buckland:2022}. The performance studies indicate that the additional 3-layer provides an improvement by a factor of 2 of the pointing resolution and of the standalone tracking efficiency down to the lowest momentum regime ($<100$~MeV/$c$). This opens a new window to explore the stages of the high-energy hadron collisions at the early stage~\cite{ITS3:2019,ALICE3:2022}. 

In recent R\&D activities, carbon foam samples have been tested in the laboratory beside carbon fiber structures. ALICE has found that carbon foams are the most promising candidates to satisfy both the low material budget and the cooling requirements, however, the detailed modelling of such a structure, involving the geometrical complexities in a finite-element simulation, is not suitable as the entire design procedure would be highly resource intensive, and it could not exclude the present uncertainties. Moreover, as such structures can show an effective non-Fourier behaviour, it is worth investigating the effects of porosity, and find a reliable way to determine the effective thermal properties. In order to obtain truthful, more precise, and resource-friendly solutions, we analyse two highly porous carbon foam samples to understand their thermal behaviour better and whether the high porosity indeed results in non-Fourier effects.

\subsection{Heat conduction in foams} 
A recent study by Lunev et al.~\cite{LunEtal22} presented the complexity of such material from both experimental and thermodynamic modeling points of view. First, they realized that the Fourier heat equation alone is not enough to properly characterize the overall transient behaviour of an aluminum foam material. In order to exclude the possibilities of any experimental artefact, they performed numerous detailed, highly resource-intensive simulations. It turned out that the measurements were acceptable, and the apparent deviation from Fourier's law is valid. It is worth emphasizing that each components of a complex structure obey Fourier's law, nonetheless, their interaction during the heat conduction process results in a non-Fourier behaviour. That is, while Fourier heat conduction is present in both the matrix and the gas phase in the inclusions, heat convection and radiation could also occur together at the same time. Consequently, they altogether result in a complex heat transfer phenomenon, for which we propose a simpler, more effective modelling approach. Additionally, the determination of effective properties for a complex structure stands as an additional challenging task, even when one deals with known microstructural material.

It is clear that such detailed simulations have enormous computational and memory requirements, needing high-end performance workstations or even clusters to conduct such computations in an acceptable time frame. While these detailed simulations can be accurate, the outcomes are valid only for a particular sample, and thus cannot be applied for any further thermal problems. Additionally, its energy demand is high, which could be a problem nowadays, this is definitely not feasible and sustainable as a standard practice.

\begin{figure}[]
	\centering
	\includegraphics[width=15cm,height=6cm]{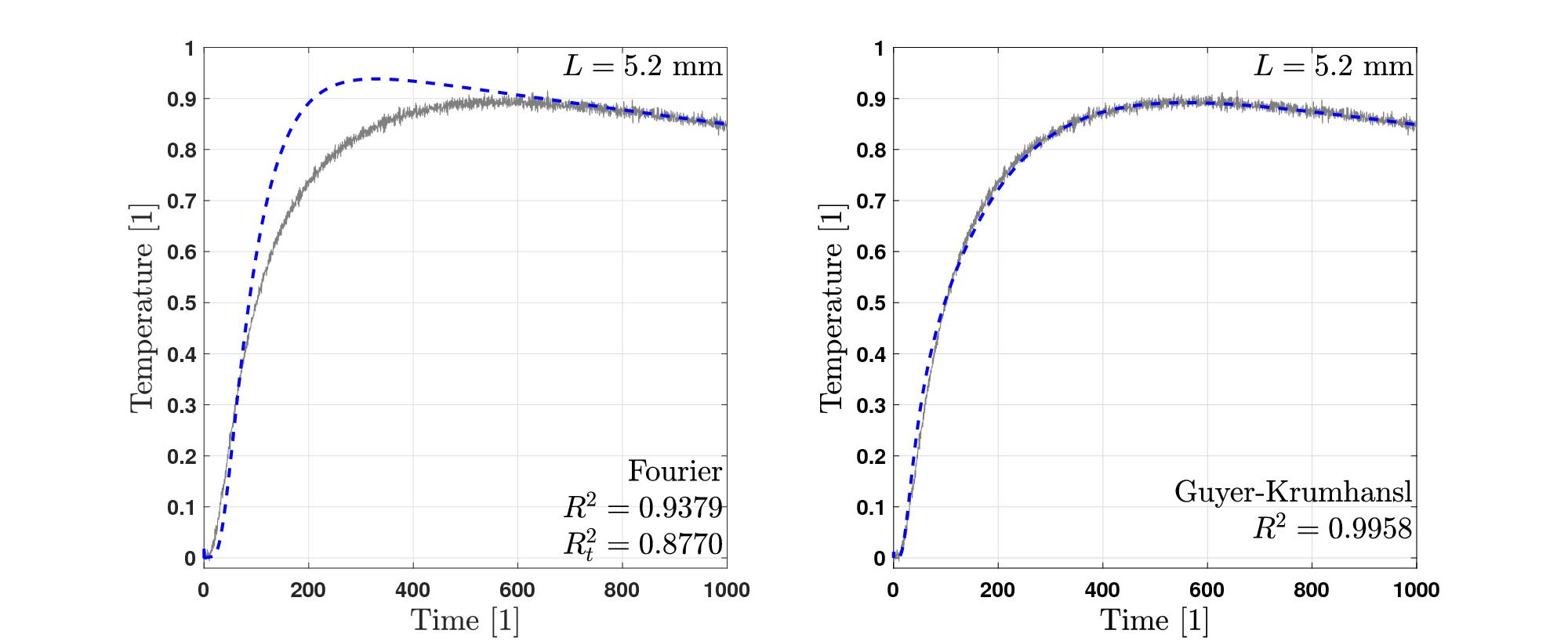}
	\caption{Typical temperature history recorded in a flash experiment on an aluminium foam sample \cite{FehKov21}. Left side: Fourier's prediction. Right side: evaluation using the Guyer\,-\,Krumhansl equation.}
	\label{fig:mf1a}
\end{figure}

Regarding the deviation from Fourier's law, we emphasize that this is characteristic on room temperature, macro-scale samples with inclusions in the order of millimeters. Such heterogeneous samples can provide an overall non-Fourier thermal response in the standard heat pulse (flash) measurement technique (see Fig.~\ref{fig:mf1a}), used to measure the thermal diffusivity of a given sample. Deviations are observable only during the transients, neither the steady-state, nor the asymptotic behaviour change. Fourier's law can lead to errors up to 30\% for thermal diffusivity and thus predicts a significantly different temperature profile. Figure \ref{fig:mf1a} presents a typical outcome in that regard, showing that Fourier's law fails to properly model the transient behaviour; first, the predicted temperature is slower, then it overshoots the measured temperature history, indicating the presence of two characteristic time scales in heat transfer \cite{FehKov21}.

We also underline the fact the representative size of such heterogeneous materials can be much larger than the limiting size of a standard laser flash apparatus or any analogous equipment.
Therefore, the sample is not necessarily representative in regard to the heterogeneity it consists, resulting in a probable size dependence~\cite{FehEtal21}. The samples we investigate here have $5$ mm thickness, which is the maximum allowable size for our equipment. Since all flash equipment suffer from such issue, it becomes inevitable to develop such a reliable evaluation procedure, which is able to capture the transient effects together with providing an effective thermal parameter. Indeed, this is a realistic scenario also for the carbon foams we aim to investigate, Bonad \cite{Bonad12} observed the strong influence of sample size. Moreover, in the Ph.D.~thesis \cite{Bonad12}, the effect of the manufacturing technology is present, which affects both the thermal conductivity and its temperature dependence. 
This is still an open question, and by exploring these difficulties, we have a step closer to the development of novel measurement techniques. However, at this moment, we can only apply the standard technique to measure the thermal diffusivity with keeping in mind the possible shortcomings. Our work extends the study of Bonad by investigating the transient behaviour of similar carbon foam samples.

The observations of Lunev et al.~are in agreement with previous experiments, which show the same deviation \cite{Botetal16, Vanetal17, FehKov21}, additionally, it is found that Guyer\,--\,Krumhansl (GK) equation is a promising candidate to model and explain the observed phenomenon, obtaining a notably better thermal description for that problem. This is a 'double-diffusive' model consisting of two characteristic time scales, therefore, it can model the interaction of different heat transfer channels (mechanisms) of foam materials by introducing two new parameters. These can be determined from a single flash experiment together with the thermal diffusivity~\cite{FehKov21}. The GK equation also provides effective parameters, similar to Fourier's law, hence no detailed, highly resource-intensive simulations are needed as the effective parameters substitute the complex inner structure. 

In the following, we provide a short overview of the experimental and modeling background we use, then we present and discuss our findings about the thermal behaviour of carbon foam samples. Furthermore, we emphasize that such highly porous carbon foams have not been investigated before and also not used as a basic structural element of such detectors. Thus it is crucial to discover, understand and offer an efficient modeling approach for the safe design. Our aim is to provide a first insight how such complex structures can effectively be modeled, study the limitations of Fourier's law and find the proper effective parameters that can be used in further simulations.

\section{Models and evaluation for heat pulse experiments} 

\subsection{Experimental arrangement} 

\begin{figure}[]
\centering
\includegraphics[width=14cm,height=6cm]{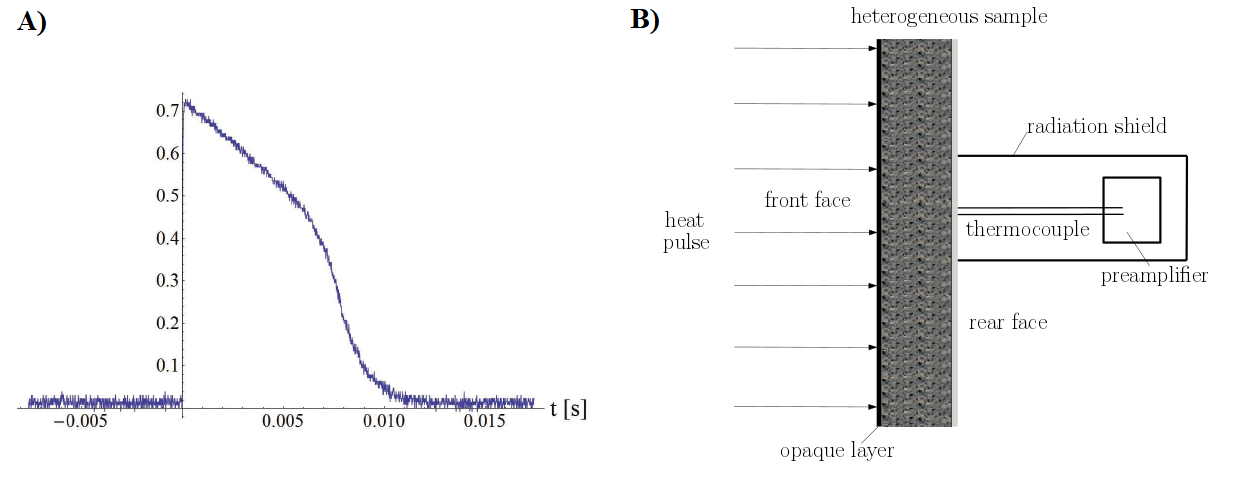}
\caption{A) Shape of the heat pulse (in arbitrary unit) \cite{Botetal16}. B) Arrangement of the experimental setup for the heat-pulse measurement of carbon foam samples.}
 \label{fig:arragement}
\end{figure}

The experiments are carried out using a well-known and used flash heat pulse method, which allows the determination of the thermal diffusivity, for our purpose,  room temperature measurements are satisfactory. The arrangement of the experiment is shown in Fig.~\ref{fig:arragement}. The excitation is provided by a flash lamp on the front face of the heterogeneous sample, while the temperature history is measured on the rear face using a K-type thermocouple. The thermocouple outputs are isolated during the measurement in order to minimize the various electrical noises and to prevent the heat pulse source from introducing any disturbance into the thermocouple circuit and, thus, into the measurement itself. An important part of the measurement is the trigger signal, the signal that shows exactly when the heat pulse occurred. This is detected and recorded by a photovoltaic sensor that induces a voltage in response to light. The measured signals are recorded with a PC oscilloscope. The advantage of the oscilloscope is that peaks due to interference can be reduced using gain settings.

\begin{figure}[]
\centering
\includegraphics[width=11cm]{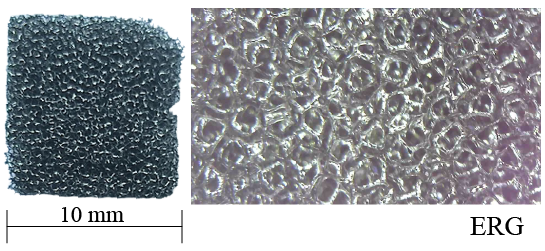}
\includegraphics[width=11cm]{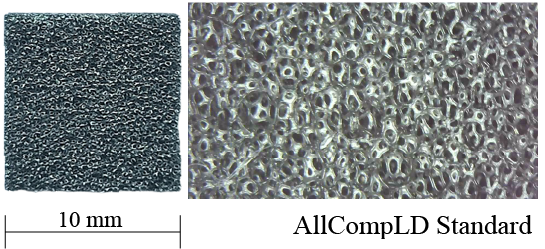}
\caption{Carbon foam samples: ERG (upper) and AllCompLD (lower) samples with 500$\times$ magnification by optical microscope.}
 \label{fig:samples}
\end{figure}

The samples used in the experiments are $10\times10\times5$~mm\textsuperscript{3} brick blocks with a random internal structure and produced by the ERG Aerospace Corporation and AllComp Inc. Two types of samples were measured, which are geometrically identical but differ in their internal structure. The higher porosity sample is called ERG Duocell~\cite{Duocell}, while the denser one is called AllCompLD K9 Hi-K Standard~\cite{AllComp}. The difference between the two samples is shown in Figure~\ref{fig:samples}. The porosity of the ERG sample is 0.97, while the porosity of the AllCompLD Standard samples varies between 0.85 and 0.9, depending on the samples.

It can be seen that both types are transparent, so measuring them without any preparation can give false results as both specimens are thermally transparent. For this reason, a sheet of graphite-impregnated paper is attached on the front side before the experiments begin. Hence the heat pulse can be absorbed on the front face while the thermal resistance of the impregnated paper has negligible impact on the overall behaviour since it has practically the same thermal conductivity as the foam matrix but with a thickness being one magnitude smaller than the sample. Otherwise, the sample would be too transparent for the thermal radiation, and the absorption would not be homogeneous on the front surface, thus such measured data cannot be interpreted. On the rear side, oppositely, we applied an aluminum foil to ensure the contact on the thermocouple. 
The graphite-impregnated paper has a thickness of $0.05$ mm, and the paper thermal conductivity is significantly increased by the graphite, up to $20$ W/(m K), therefore its thermal resistance can be found in the order of $10^{-5}$ m$^2$K/W. The aluminum foil has even less thermal resistance, it is about $10^{-7}$ m$^2$K/W due to the large thermal conductivity and the thickness of $0.02$ mm. The sample, on the other hand, has a thickness of $5$ mm, and let us assume that its effective thermal conductivity ($\lambda_{\textrm{eff}}$) can be estimated with 
\begin{align}
 \lambda_{\textrm{eff}} = (1-\phi)\lambda_{\textrm{carbon}} + \phi \lambda_{\textrm{air}}
\end{align}
with $\phi=0.85$ porosity. Although this is a rough estimate, which we will revisit later, it shows that the effective thermal conductivity is about $3$ W/(m K), therefore the sample thermal resistance is in the order of $10^{-3}$ m$^2$K/W. Overall, both the impregnated paper and foil thermal resistances are two orders of magnitude smaller than that of the investigated specimen. Consequently, we assume that these additional parts do not significantly distort the effective thermal diffusivity values, which we will find from the fittings. We also admit that the contact resistances between the graphite covering, sample, and foil, can still be a factor, which we tried to reduce as much as possible with spring-loaded cantilevers on the edge of the sample. While this does not completely ensure that the contact resistances are minimized, one must also consider the fact that the contact surface of such strongly porous sample is also small, and most probably this is not the crucial element in the overall heat transfer.

\subsection{Heat Conduction models}

The first building block is the balance equation of initial energy represented, in which $e=cT$ with $c$ being the isochoric specific heat, $\rho$ is the mass density, and reads as
\begin{align}
\rho c \partial_t T + \partial_x q = 0, \label{eqv:bale}
\end{align}
for heat conduction in solid bodies without heat sources and mechanical coupling, hence the sample is considered to be rigid. We also assume a one-dimensional model to be adequate to evaluate the measured temperature history, so that the divergence of the heat flux $q$ reduces to $\partial_x q$; and $\partial_t$ denotes the partial time derivative. Furthermore, all coefficients are assumed to be constant and independent of temperature due to the small temperature increment ($3-5$ K) during the measurement. 

In order to mathematically and physically close the balance equation of energy (\ref{eqv:bale}), one needs to consider a so-called constitutive equation in accordance with the second law of thermodynamics. The usual candidate is Fourier's law,
\begin{align}
 q = - \lambda \partial_x T,
\end{align}
which fails to accurately describe many of the room temperature measurements performed on heterogeneous structures such as rocks and foams \cite{Botetal16, FehEtal21}. In these cases, the Guyer\,--\,Krumhansl equation was found to be the least necessary extension of Fourier's law \cite{FehKov21}, which reads in one spatial dimension
\begin{align}
\tau_q \partial_t q + q+ \lambda \partial_x T - \kappa^2 \partial_{xx} q=0. \label{gk1}
\end{align}
Here, $\tau_q$ is the relaxation time for the heat flux $q$, and $\kappa^2$ is a sort of intrinsic length scale. Whereas it was first derived on the basis of kinetic theory \cite{GuyKru66a1}, this model also has a strong background in non-equilibrium thermodynamics with internal variables (NET-IV), in which the new coefficients originate from the Onsager relations and are restricted by the second law only \cite{Van01a, VanFul12}. Moreover, since the derivation of the GK equation exploits the energy balance (\ref{eqv:bale}) as a constraint, (\ref{eqv:bale}) is naturally satisfied. It is crucial to emphasize that the continuum GK equation \eqref{gk1} is free from any assumption on the microscopic mechanisms, therefore this model is free from the usual limitations of kinetic theory, and it is applicable for room temperature problems with very low Knudsen number, too. Consequently, although the structure of Eq.~\eqref{gk1} completely identical with the Guyer-Krumhansl equation, its interpretation is different. Indeed, it can be understood as a double-diffusive model, since the ratio $\kappa^2/\tau_q$ is also a diffusivity-like quantity, and comparable with the thermal diffusivity $\alpha=\lambda/(\rho c)$. The experimentally crucial domain is characterized by $\kappa^2\tau_q > \alpha$, this is called over-diffusion \cite{Botetal16, FehEtal21}, requiring a heat equation at least two conduction time scales.
The data recorded with a heat pulse experiment is proved to be eligible to find all the parameters appearing in Eq.~\eqref{gk1}, and these parameters effectively characterize the heterogeneous material structure. 

It is more convenient to utilize the dimensionless form of the governing equations,
\begin{subequations}\label{eqv}
\begin{equation}\label{eqv_b}
    \partial_{\hat{t}} \hat T + \partial_{\hat{x}} \hat q=0,
\end{equation}
\begin{equation}\label{eqv_f}
  \hat  q + \hat \alpha \partial_{\hat{x}}\hat T = 0,
\end{equation}
\begin{equation}\label{eqv_gk}
   \hat {\tau}_q \partial_{\hat{t}}\hat q +\hat q + \hat\alpha \partial_{\hat{x}} \hat T -\hat\kappa^2\partial_{\hat{x}\hat{x}}\hat q = 0.
\end{equation}
\end{subequations}
together with the parameters defined in Table \ref{table:DL_params}, following \cite{Botetal16, FehKov21}. In what follows, we neglect the hat notation and we use dimensionless parameters by default, otherwise we explicitly denote the units if necessary. 

We wish to emphasize the fundamental difference between the Fourier and GK heat equations
with presenting the corresponding analytical solutions for the present experimental setting. For the detailed calculation, we refer to the Appendix in \cite{FehKov21}. In classical heat transfer, it is usual to utilize a so-called one-term solution, viz., only the first of the infinite series is used to model a temperature history. Naturally, such approach is limited, and cannot be used for the initial transients. For the rear side temperature history, the Fourier and GK solutions are 
\begin{gather}
	T(x{=}1,t) = Y_0 \exp\left( -h t \right) - Y_1 \exp\left( x_F t \right),\quad \textnormal{with} \quad  x_F=-2h -\alpha \pi^2 <0, \label{a1}
\end{gather}
\begin{gather}
	T(x{=}1,t) = Y_0 \exp\left( -h t \right) - Z_1 \exp\left( x_1 t \right) - Z_2 \exp\left( x_2 t \right),\quad \textnormal{with} \quad x_1, x_2 < 0, \label{a2}
\end{gather}
in which $x_F$, $x_1$, and $x_2$ are proportional with the corresponding conduction time scales, all are influenced by the heat transfer coefficient $h$. Compared to the Fourier solution, the GK equation consists of two conduction time scales represented by $x_1$ and $x_2$, and $x_2<x_F<x_1$ holds.

Recently, an advanced evaluation procedure was developed to estimate the GK parameters immediately, based on an analytical solution \cite{FehKov21}. Here, we apply an improved version, including an iterative procedure based on particular sensitivity functions. During the iteration, we implement these analytical solutions, however, with $100$ terms from the infinite series, i.e., this becomes free from the limitations of the one-term solution.

\begin{table}[h!]
\centering
\begin{tabular}{ l c c c} 
\textrm{Time and {spatial coordinates}:} & $\hat{t} =\frac{t}{t_p}$ &  \textrm{and}  & $\hat{x}=\frac{x}{L}$, \\
\textrm{Thermal diffusivity:} &  $\hat \alpha = \frac{\alpha t_p}{L^2}$ & \textrm{with} & $\alpha=\frac{\lambda}{\rho c}$, \\
\textrm{GK parameters:} &  $\hat{\tau}_q =\frac{ \tau_q}{t_p}$ \quad & \textrm{and} \quad &  $\hat{\kappa}^2 = \frac{\kappa^2}{L^2}$, \\
\textrm{Temperature:} & $\hat{T}=\frac{T-T_{0}}{T_{\textrm{end}}-T_{0}}$ \quad & \textrm{with} \quad & $T_{\textrm{end}}=T_{0}+\frac{\bar{q}_0 t_p}{\rho c L}$, \\
\textrm{Heat flux:} & $\hat{q}=\frac{q}{\bar{q}_0}$ \quad & \textrm{with} \quad
& $\bar{q}_0=\frac{1}{t_p}  \int_{0}^{t_p} q_{0}(t)\textrm{d}t$, \\
\textrm{Heat transfer coefficient:} & $\hat h= h \frac{t_p}{\rho c}$, & & \\
\textrm{Fourier resonance condition:} & $\hat \kappa^2/\hat \tau_q = \hat \alpha$. & & \\
\end{tabular}

\caption{Dimensionless quantities following \cite{Botetal16, FehKov21}, where $t_p$ denotes the heat pulse duration for which interval $\bar q_0$ averages the heat transferred by the heat pulse. Here, $L$ is the sample thickness. $T_{\textrm{end}}$ represents the adiabatic steady-state, and $T_0$ is the uniform initial temperature.}
\label{table:DL_params}
\end{table}

\subsection{Evaluation of heat pulse measurements}
In a heat pulse experiment, one obtains the rear side temperature as a function of time, and that temperature history is used to determine the unknown thermal parameters. For the Fourier heat equation, these unknown parameters are the heat transfer coefficient $h$ and the thermal diffusivity $\alpha$. For the GK equation, that set of parameters is extended with $\tau_q$ and $\kappa^2$. 
The iteration is based on the so-called local sensitivity functions, i.e.
\begin{align}
	S_{p_i,t} = \frac{\partial y}{\partial p_i}, \quad y=y(p_i, t), \quad i=1,\dots,N,
\end{align}
in which $y(p_i, t)$ denotes the time series predicted by the model, in our case, this is 
the rear side temperature, and that depends on $N$ number of parameters $p_i$. Thus, the sensitivity function $S_{p_i}$ characterizes how a the temperature history is influenced by changing a given parameter in the model. Furthermore, this is called local sensitivity, since the model nonlinearly depend on the unknown parameters, and therefore, the sensitivity function can remarkably change depending on the value of the parameter. We determine the corresponding sensitivity functions for each step of iteration.
Although with the analytical solution the sensitivity function $S_{p_i}$ could be expressed analytically, it cannot be done in a reasonably simple form, hence we approximate each sensitivity function with
\begin{align}
	S_{p_i,t} = \frac{\partial y}{\partial p_i} \approx \frac{T(p_i+\Delta p_i,t)   -T(p_i,t)}{\Delta p_i}, \quad  \Delta p_i = 0.05 p_i, \label{eq4}
\end{align}
in which we approximate the partial derivative similarly to finite differences. Additonally, in order to determine $\Delta p_i$, we impose a 5\% difference for a given parameter value.

After determining the set of sensitivity functions for each parameters, one can construct a so-called sensitivity matrix $\mathbf S$, such that $\mathbf S = [S_{p_1} S_{p_2} \dots S_{p_N}]$. Now, let $P_k$ stand for the $k$th iteration of the parameters, hence
\begin{align}
	P_{k+1} = P_k + (\mathbf S_k^T \mathbf S_k)^{-1} \mathbf S^T_k (T_{\textrm{measured}} - T (P_k)),
\end{align}
iteration formula determines the unknown parameter, for which the sensitivity functions consist of the model attributes as well. This converges only if all columns of the matrix  $\mathbf S$ are linearly independent, i.e., all parameters are independent. Otherwise, the matrix $(\mathbf S_k^T \mathbf S_k)$ becomes singular, and that is an immediate indication of 
the presence of redundant parameters in the model. For the GK equation, that happens only when the so-called Fourier resonance $\alpha = \kappa^2/\tau_q$ holds, as the solution of the GK equation reduces to Fourier, and thus the model parameters are not independent. Since we always test first the Fourier heat equation, Fourier resonance does not influence the effectiveness of the iteration procedure.

\section{Experimental findings}\label{experiement} 
We thoroughly investigated the obtained measurement data on the ERG and AllCompLD samples. Both models, Fourier and GK, have been applied on the measured data. Here we summarize the obtained result.

\subsection*{ERG samples}  
\begin{figure}[H]
\centering
\includegraphics[width=11cm]{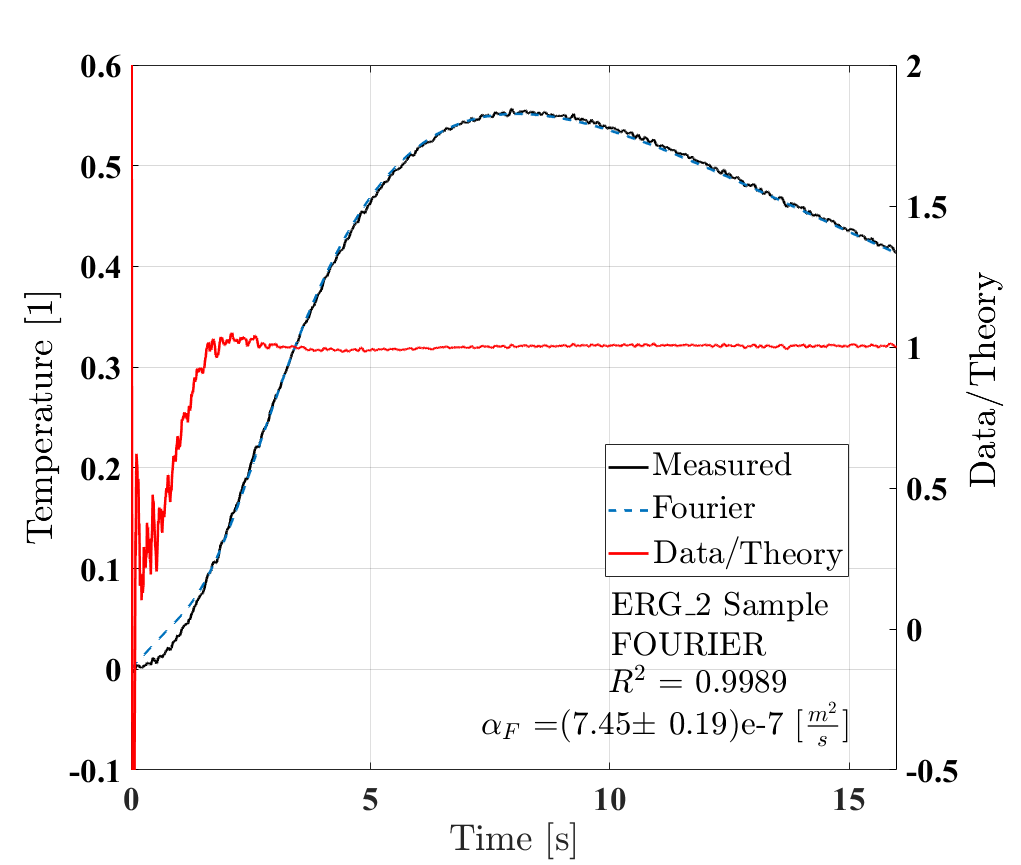}
\caption{ERG\_2 sample evaluation with the Fourier theory.}
 \label{fig:ERG_F}
\end{figure}

It is found that the ERG sample with a looser structure does not show any non-Fourier phenomenon, Fourier equation is proved to be appropriate for all the three samples we received and measured. The corresponding thermal diffusivity values are summarized in Table~\ref{tab:ERG}, found as an average of multiple measurements on three different ERG samples. The notable difference between ERG\_1 and the others possibly originates from the different structures. Despite that the samples have the same porosity level, the structure itself possesses a statistical variation, which can lead to deviating values for each samples. For an industrial purpose, such measurement would require a much larger series of samples for which the present statistical variations can be excluded overall.
Figure~\ref{fig:ERG_F} shows a typical outcome for an ERG sample. Small deviations are present right after the trigger pulse, however, this is not notably influential and does not distort the fitting procedure.
\begin{table}[H]
\centering
\begin{tabular}{cc}

Sample ID & $\alpha_F$                       \\ 
          & $10^{-7}$ {[}$\mathrm{m^2}$/s{]} \\ \hline
ERG\_1    & 6.06    $\pm$ 0.12                         \\ \hline
ERG\_2    & 7.45 $\pm$ 0.19                            \\ \hline
ERG\_3    & 7.37 $\pm$ 0.54                                                       
\end{tabular} 
\caption{Thermal diffusivity of the ERG samples by the Fourier equation.}
 \label{tab:ERG}
\end{table}

\subsection*{AllCompLD samples} In contrast to the ERG sample, the more dense AllCompLD Standard sample showed non-Fourier behaviour for all samples we received. The found thermal parameters are summarized in Table \ref{tab:AllComp}. 

\begin{table}[H]
\centering
\begin{tabular}{cccccc}
Sample ID  & $\alpha_F$                       & $\alpha_{GK}$                      & $\tau_q$ & $\kappa^2$    &      Calculated $\alpha_F$             \\
           & $10^{-6}$ {[}$\mathrm{m^2}$/s{]} & $10^{-6}$ {[}$\mathrm{m^2}$/s{]} & {[}s{]}  & $10^{-6}$ {[}$\mathrm{m^2}${]} & $10^{-6}$ {[}$\mathrm{m^2}$/s{]}  \\ \hline
AllComp\_1 & 8.25 $\pm$ 0.74                             & 6.9 $\pm$ 0.45                              & 0.160 $\pm$ 0.005    & 2.432 $\pm$ 0.096       &11                   \\ \hline
AllComp\_2 & 5.78 $\pm$ 0.27                             & 4.55 $\pm$ 0.2                             & 0.207 $\pm$ 0.032    & 2.385 $\pm$ 0.167 & 8.03   \\

\end{tabular}
\caption{Thermal diffusivity of the AllCompLD samples.}
 \label{tab:AllComp}
\end{table}

\begin{figure}[H]
\centering
\includegraphics[width=18cm]{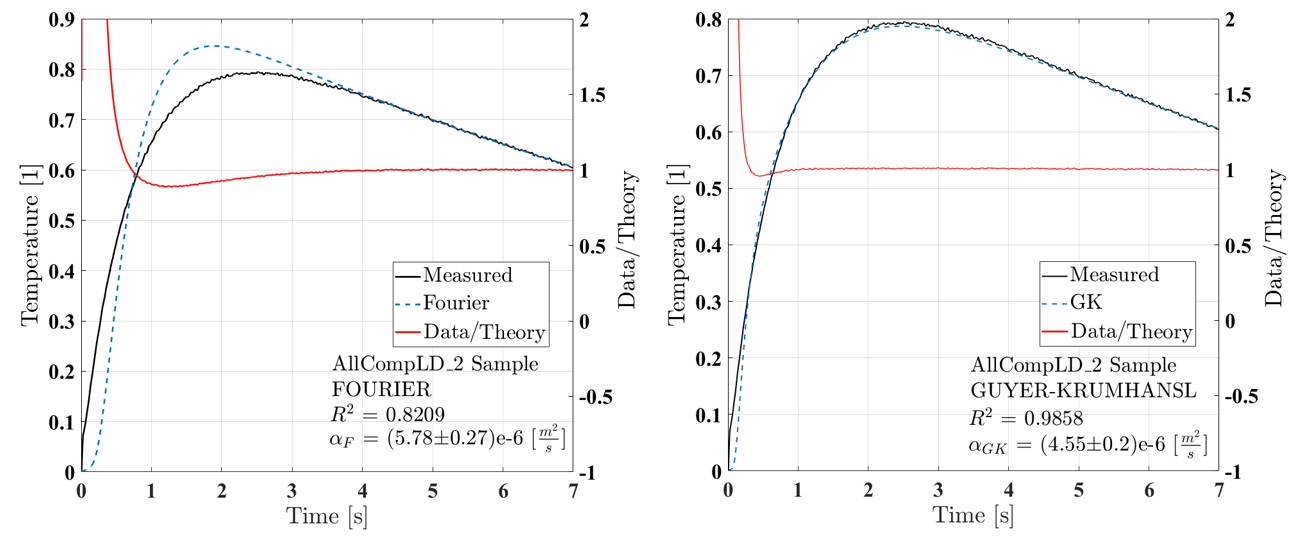}
\caption{AllCompLD\_2 sample evaluation with the Fourier and GK heat equation.}
 \label{fig:AllComp_F}
\end{figure}

The outcome of the AllCompLD\_2 sample measurements is shown in Figures~\ref{fig:AllComp_F}. It can be clearly seen that the Fourier fitting was not successful, and that model cannot be fitted to the measurement data (Figure~\ref{fig:AllComp_F}), producing the same type of deviation that was experienced before in \cite{Vanetal17}. In such a case, at the beginning of the measurement, Fourier's prediction is slower than the actual temperature measured. Approaching the top, the predicted temperature signal becomes faster than the measured one. After that, typically after $4$ s, cooling is dominating the process, the heat transfer becomes slow enough, and therefore both curves run together.
It can be seen that the fit at the beginning of the measurement is also different in the GK case, this may be due to the transparency for thermal radiation, which we could not fully eliminate during the measurements. This is natural since the sample is highly porous and inevitable transparent.

It can be observed that the random internal structure of the samples also strongly influences the thermal diffusivity so that at such small sizes, such number of samples cannot provide an entirely reliable value for each parameter. This is not surprising, and this is also not our main objective here. Let us focus about the effective parameters, and on further possibilities.

Since the transient region cannot be modeled with the Fourier heat equation, the resulting Fourier thermal diffusivity $\alpha_F$ cannot be reliable in any way. This situation proves the advantage of a more detailed model such as the GK equation. The additional parameters are helpful to derive an effective thermal diffusivity, this is represented by 'Calculated $\alpha_F$' in Table \ref{tab:AllComp}. This can be achieved by exploiting
\begin{align}
 \alpha_F \approx \frac{1}{2} \left( \alpha_{GK} + \frac{\kappa^2}{\tau_q}\right), \label{average}
\end{align}
first observed experimentally on several rock samples \cite{FehEtal21}. Eq.~\eqref{average} expresses that the effects of multiple heat transfer channels can be averaged for a long-time behaviour, and that averaging yields an effective thermal diffusivity for the Fourier heat equation. In other words, the utilization of the GK equation can offer two advantages. First, it provides an efficient and a more detailed model, which is able to describe the fast transients. Second, if the fast initial transients are less important in a practical application, especially for relatively slow heat transfer processes, then the GK equation offers a proper effective thermal diffusivity for the Fourier equation. Therefore, one can exploit the outcome of the GK equation without directly implementing such a complex model. In the following, let us investigate further details of the effective parameters.

\section{Effective thermal conductivity}
While the heat pulse experiments can provide an effective thermal diffusivity, especially with the use of the GK heat equation, thermal conductivity is still needed in many practical applications. The ERG and AllComp samples are partially characterized by their producers \cite{Duocell, AllComp}.
According to the available information \cite{Duocell, AllComp}, for the ERG samples, the product sheet indicates $\lambda = 0.033 - 0.05$ W/(m K). Regarding the AllComp foams, the thermal conductivity is  $\lambda = 20 - 40$ W/(m K), which contribute to the observed order of magnitude difference in the corresponding thermal diffusivity. For the mass density, the ERG sample is extremely lightweight, $\rho \approx 70$ kg/m$^3$, while the AllComp samples have $\rho=220$ kg/m$^3$.
For the specific heat, the ERG sheet indicates only a single value of $1260$ J/(kg K), and it must be measured for the AllComp sample. 

\begin{figure}[H]
	\centering
	\includegraphics[width=11cm]{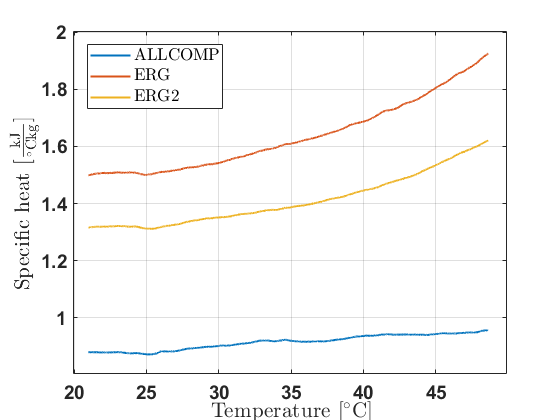}
	\caption{Specific heat measurement results of the studied carbon matrices.}
	\label{fig:fajho}
\end{figure}

Therefore, in order to achieve a more complete understanding, we performed specific heat measurements. The specific heat of the samples was measured using a Perkin-Elmer DSC 8000, which is suitable for measuring dust samples, so the porosity of the samples does not appear here as a characteristic quantity. The measurements were preceded by precision mass measurements, with sample weights ranging from 3-5 mg. The measuring temperature range was between 20-50 $^\circ$C since the specific heat values measured at this temperature range are relevant for the detector application. Figure \ref{fig:fajho} clearly shows that although all samples are made from carbon, there is a large variation between individual samples, with the AllComp sample having a specific heat of almost half the value of the other samples. While the AllComp sample shows a slight change with respect to temperature, the ERG and ERG2 samples show a more considerable dependence. Compared to the initial temperature, the AllComp sample increased by 8.6 $\%$, the ERG sample by 28.3 $\%$, and the ERG2 sample by 23.18 $\%$.

Furthermore, the characteristic values and the corresponding temperature-dependences confirm that the foam matrices are made from different forms of carbon. The ERG sample is made of amorphous carbon with thermal conductivity of $1$ W/(m K). On contrary, the AllComp uses polycrystalline graphite matrix with thermal conductivity of $80$ W/(m K) on room temperature.
On the one hand, it validates our initial assumption about the order of the thermal resistance of the samples. On the other hand, it provides an insight into over-diffusion. That propagation mode becomes essential for considerable differences between the two heat transfer channels. For the highly porous ERG samples, the low volume ratio of the carbon matrix cannot induce strong over-diffusion effects. The AllComp sample, however, has lower porosity (85\% instead of 97\%), thus the matrix material becomes more dominant in the heat transfer process, and that dominance was eligible to observe over-diffusion.

The measured specific heat values are characteristic for the carbon matrix, and naturally differ from the effective properties. To obtain the corresponding effective values, we use
\begin{align}
 c_{\textrm{eff}} = (1-\phi) c_{\textrm{carbon}} + \phi c_{\textrm{air}}, 
\end{align}
therefore, it yields $c_1 = 1016$ J/(kg K) for the first, and $c_2=1011$ J/(kg K) for the second and third ERG samples using 97 \% porosity. For the AllComp with 85 \% porosity, we obtain $911$ J/(kg K).  Table \ref{tab:properties} presents the calculated values of thermal conductivity, where  $\alpha_{Fc}$ is obtained from the averaging the GK coefficients based on \eqref{average}, and the subscript "c" aims to emphasise it. Consequently, $\lambda_{Fc}$ follows from $\alpha_{Fc}$. 

\begin{table}[H]
	\begin{tabular}{ccccccc}
		& $\rho$                 & $c$                    & $\alpha_{Fc}$                   & $\alpha_{GK}$                   & $\lambda_{Fc}$       & $\lambda_{GK}$       \\
		\multirow{-2}{*}{Sample ID} & {[}$\frac{kg}{m^3}${]} & {[}$\frac{J}{kg K}${]} & $10^{-6}$ {[}$\frac{m^2}{s}${]} & $10^{-6}$ {[}$\frac{m^2}{s}${]} & {[}$\frac{W}{mK}${]} & {[}$\frac{W}{mK}${]} \\ \hline
		ERG\_1                      & 70                     & 1016                   & 0.606   & -       & 0.043                & -                    \\ \hline
		ERG\_2                      & 70                     & 1011                   & 0.745                           &-       & 0.052                & -                    \\ \hline
		ERG\_3                      & 70                     & 1011                   & 0.737                           & -       & 0.052                & -                    \\ \hline
		AllComp\_1                  & 220                    & 911                    & 11                              & 6.9     & 2.207                & 1.383                \\ \hline
		AllComp\_2                  & 220                    & 911                    & 8.03                            & 4.55    & 1.610                & 0.911               
	\end{tabular}
\caption{Summary of sample properties.}
\label{tab:properties}
\end{table}

The effective thermal conductivity can be estimated in various ways, three of which are mentioned below, only for demonstrative purposes. We use the following estimations:

Voigt-type:
\begin{align}
	\lambda_{eff} = V_1 \lambda_1 + V_2 \lambda_2 ;
\end{align}

Markworth et al.: 
\begin{align}
	\lambda_{eff} = V_1 \lambda_1 + V_2 \lambda_2 + V_1 V_2 \frac{\lambda_1-\lambda_2}{\frac{3}{\frac{\lambda_2}{\lambda_1}-1}+V_1};
\end{align}

Wakashima-Tsukamoto:
\begin{align}
	\lambda_{eff} = \lambda_1 + \frac{\lambda_1 V_2 (\lambda_2-\lambda_1)}{\lambda_1 + \frac{(\lambda_1-\lambda_2) V_1}{3}};
\end{align}
where $V_1$ and $V_2 = 1 - V_1$ are the volume ratios, and $\lambda_1$ and $\lambda_2$ are the thermal conductivity of the components. The effective thermal conductivities calculated for the different types are given in Table \ref{tab:effectiv_calc}.

In regard to the ERG samples, both the Voigt-type estimate and the measurements are in accordance with the data sheet provided by the manufacturer. The other formulae underestimate the thermal conductivity, but their outcomes remain in the same order of magnitude. These results also indicate that our measurement procedure does not distort the thermal diffusivity values. 

For the AllComp samples with 85\% porosity, however, the situation is different. While the data sheet provided $20-40$ W/(m K), both the estimates and the measurements differ significantly. Although our measurements show an indeed notably higher thermal conductivity compared to the ERG samples, these values are much lower than that from the data sheet. The closest value is found by the Voigt-type estimate, the other formulae are definitely of no use for carbon foams. The observed difference can occur due to the sample size: thermal conductivity measurements use notably larger samples for which the local variations can vanish. The second reason could be a possible anisotropy: for smaller samples such as we used in our measurements, the local structural variations can cause that sort of deviation in thermal conductivity. However, all heat pulse equipment suffer from the exact same issue, the sample size is strongly limited and falls far from the representative sample size corresponding for the given heterogeneous structure. 

Nevertheless, the Guyer-Krumhansl heat equation provided a considerably better transient description, applicable without restrictions on the time scales. If one wishes to keep the Fourier heat equation, $\alpha_{Fc}$ can be a more suitable thermal diffusivity to model relatively slow heat transfer processes. In the present situation, it means that with a $0.01$ s long heat pulse, the Fourier model will be applicable after $\approx 3$ s. The present investigation highlight that such highly porous samples can bear size-dependent behaviour, that vanishes for notably larger samples. However, this is not necessarily applicable in the practice, the designed foam structure in the CERN-ITS detectors could have a thickness between $\approx 5-10$ mm, so the practical utilization is also a constraint.

\begin{table}[H]
	\begin{tabular}{cccc}
		\multirow{2}{*}{Sample ID} & \multicolumn{3}{c}{$\lambda_{eff} [\frac{W}{mK}]$} \\ \cline{2-4}
		& Voigt-type   & Markworth   & Wakashima-Tsukamoto   \\ \hline
		ERG                        & 0.055        & 0.028       & 0.023                 \\ \hline
		Allcomp                    & 12.02        & 0.039       & 0.012                
	\end{tabular}
\caption{Summary of calculated effective thermal conductivity.}
\label{tab:effectiv_calc}
\end{table}

\section{Discussion} 

The heat pulse (flash) experiment has been performed on several novel carbon foam samples for the first time. Two samples with porosity of 0.97 (ERG) and 0.85-0.90 (AllCompLD) by the ERG Aerospace Corporation and AllComp Inc.~has been measured, respectively.  The flash method is a well-known and widely accepted procedure to measure the thermal diffusivity, however, for foams, the found thermal diffusivity can greatly depend on the structure and on the possible non-Fourier behaviour. We prepared the samples with graphite-coated, impregnated paper to achieve the homogeneous boundary condition for all measurements. The measured data were fitted with the Fourier and the Guyer\,--\,Krumhansl heat conduction equations.

First, we obtained that ERG samples follow the Fourier model well, and provide a thermal diffusivity parameter, $\alpha_F=(6.96 \pm 1.02) \times 10^{-7}$~m\textsuperscript{2}/s ($R^2=0.9989$), indeed the Guyer\,--\,Krumhansl solution for this sample results in the Fourier limit as well.
On the contrary, AllCompLD samples showed non-Fourier behaviour, which can significantly detune the thermal diffusivity. Moreover, it necessitates the use of additional parameters, too, which all can be determined using the same temperature history.
For Fourier fits, $\alpha_F=(7.02 \pm 1.7) \times 10^{-6}$~m\textsuperscript{2}/s ($R^2=0.8209$). Moreover, the measured data was in good agreement with the Guyer\,--\,Krumhansl model with parameters,  $\alpha_{GK}=(5.73 \pm 1.5) \times 10^{-6}$~m\textsuperscript{2}/s, $\tau_q= (0.1841 \pm 0.055) $~s, and $\kappa^2= (2.41 \pm 0.121) \times 10^{-6}$~m\textsuperscript{2}. We also proposed to calculate the Fourier thermal diffusivity based on the GK parameters using Eq.~\eqref{average} since the deviation makes impossible to properly and reliably evaluate the recorder temperature history with Fourier's law.

Following these results, we conclude that finite element calculations can safely be carried out for the ERG sample during the design of the detector. Moreover, it is now possible to perform the necessary simulations effectively, i.e., neglecting the detailed structure and substituting it with a homogeneous one. This saves a huge amount of energy and effort and greatly decreases computational costs.
Regarding the AllCompLD samples, the situation is more challenging due to the non-Fourier behaviour, and the implementation of the Guyer\,-\,Krumhansl equation for a finite element environment is still not straightforward. Adding that the parameters show significant porosity dependence, more thorough experimental and theoretical research is needed, e.g., by investigating a wide range of samples with varying porosity levels. The authors are about to continue the measurements with further types and different porosity carbon foam samples, also paying particular attention to the possible anisotropic properties. 

\section{Summary}
Determining the thermal material parameters of the carbon foams under investigation is a challenging task with uncertain outcomes. That uncertainty is present for both steady and transient measurement methods. The data sheet provided by the manufacturer of the present carbon foam samples also consisting thermal conductivity values within a relatively large interval. In other words, as the thermal diffusivity is directly proportional to the thermal conductivity, that uncertainty is inherited from the diffusivity measurements, and the resulting difference is clearly visible between the foam samples, despite the similar porosity level. Furthermore, there is also a surprising difference between the specific heat of the samples, and these measurements emphasize the need to discover the temperature dependence of thermal conductivity, too.

The values of the measured thermal parameters also have a significant variance, such as the thermal diffusivity, heat transfer coefficient, and specific heat. Measurement is challenging on small sample sizes, as the equipment itself places strong size limits on the samples, and for such high porosity objects, it is difficult to determine reliable parameters. 
As it is visible from Figures \ref {fig:ERG_F} and \ref{fig:AllComp_F}, the initial time interval is influenced by radiation penetration, however, its effect vanishes quickly and is not significantly influential for the entire measurement as we utilized graphite-coated paper for better absorption. The measured deviation from Fourier's law provides the necessary time scale, which characterizes the validity limit and beyond that, the classical heat equation seems to be an appropriate choice. We also conclude that further investigations are necessary, including larger samples with a different measurement technique.
With a larger sample volume, transient effects can be superimposed, reflecting the actual properties of the whole material. However, at the point of use, it is not always possible to incorporate a large size and volume of material, so it is a major engineering challenge to find the proper material parameters for the practice.

\section{Acknowledgement}
The authors are thankful to Ildiko Pethes and László Temleitner for helping with the specific heat measurements. 

The research reported in this paper and carried out at BME has been supported by the grants National Research, Development and Innovation Office-NKFIH FK 134277, K 135515, 2021-4.1.2-NEMZ\_KI-2021-00009, and by the NRDIO (NKFIH) Fund (TKP2020 NC, Grant No. BME-NC) based on the charter of bolster issued by the NRDI (NKFIH) Office under the auspices of the Ministry for Innovation and Technology. This paper was supported by the János Bolyai Research Scholarship of the Hungarian Academy of Sciences (KR). Samples were provided by C. Gargiulo and the ALICE Experimental Collaboration (CERN).

\bibliographystyle{unsrt}

\end{document}